# On the behavior of scattering phases in collisions of electrons with multi-atomic objects

M. Ya. Amusia[1;2] and L. V. Chernysheva[2]

[1]*The Racah Institute of Physics, the Hebrew University of Jerusalem, Jerusalem 91904, Israel*
[2]*A. F. Ioffe Physical-Technical Institute, St. Petersburg 194021, Russian Federation*

**Abstract:**
   We have studied the energy dependence of several first scattering phases with multi-atomic object. As concrete examples representing the general trends endohedrals Ne@$C_{60}$ and Ar@$C_{60}$ are considered. It appeared that the presence of an inner atom, either Ne or Ar, qualitatively affects the scattering phases, in spite of the fact that the fullerene consists of 60 carbon atoms, while the atom staffed inside is only one. Calculations are performed in the one-electron Hartree-Fock (HF) and random phase approximation with exchange (RPAE) for the inner atom while the fullerenes shell is substituted by static potential without and with the polarization potential. It appeared that the total endohedral scattering phase is simply a sum of atomic, Ne or Ar, and fullerenes $C_{60}$ phases, contrary to the intuitive assumption that the total phases on $C_{60}$ and Ne@$C_{60}$ or Ar@$C_{60}$ has to be the same.

**1**. We suggest after performing calculation of a concrete example that the total partial wave $l$ phase $\delta_l^{A@C_N}$ of an electron scattered upon endohedral A@$C_N$ is with good accuracy equal to the sum of scattering phases $\delta_l^A$ and $\delta_l^{C_N}$ of electrons upon atom A that is stuffed inside the fullerene $C_N$. We generalize this result to any target that consists of loosely bound atoms. This result is counter-intuitive, since it is reasonable to assume, at least in the case of the considered target A@$C_N$ that the scattering is absolutely dominated by the very "big atom" [1] so that the contribution of atom A can be neglected.

   The general properties of the behavior of a scattering phase upon a static potential $U$ is well established and described in text books (see e.g. [2]). It is demonstrated that if the phases $\delta_l(E)$ as functions of energy $E$ are normalized in such a way that $\delta_l(E \to \infty) \to 0$, one has $\delta_l(0) = n_l \pi$, where the number of bound electron states with angular momentum $l$ in the system $e + U$ is $n_l$.

   If the target consists of electrons and nuclei and exchange between incoming and target electrons is taken into account, another expression takes place $\delta_l(0) = (n_l + q_l)\pi$, where $q_l$ is the number of bound electron states with the angular momentum $l$ in the target itself [3]. Therefore, the behavior of phases as functions of $E$ is qualitatively different in cases when we treat the target in Hartree or Hartree-Fock (HF) approximation. In these two cases the phases deviate from each other (in numbers of $\pi$) although the strength of the potential is almost the same.

   To clarify the role of a single atom embedded in the multi-atomic system, we consider in this Letter electron scattering on concrete endohedrals Ne@$C_{60}$ and Ar@$C_{60}$, which are fullerene $C_{60}$, stuffed by a Ne or Ar atom. It is known that both of them are located at the $C_{60}$ center that considerably simplifies the calculations. The Ne and Ar atom are treated in the HF approximation and RPAE frames, while $C_{60}$ is represented by a static square well potential $W_C(r)$, which parameters are chosen to represent the experimentally known electron affinity of $C_{60}^-$ and low- and medium energy photoionization cross-sections of $C_{60}$ [4]. Fullerenes are highly polarizable objects. This is why along with $W_C(r)$ polarization potential $V_C^{pol}(r)$ has to be taken into account.

**2.** In order to obtain scattering phases for an atom, one has to solve numerically the HF equations for radial parts of the one-electron wave functions $P_{El}^A(r)$. For spherical atoms they can be presented as[1]

$$\left( \frac{1}{2}\frac{d^2}{dr^2} + \frac{Z}{r} - \hat{V}_{HF}(r) - \frac{l(l+1)}{2r^2} - E \right) P_{El}^A(r) = 0, \tag{1}$$

where Z is the inner atom nuclear charge and $\hat{V}_{HF}(r)$ is the operator of HF non-local potential (see definition in e.g. [5]). The scattering phase is determined by the asymptotic of $P_{El}^A(r)$

$$P_{El}^A(r)|_{r \to \infty} \approx \frac{1}{\sqrt{\pi p}} \sin\left[ pr - \frac{\pi l}{2} + \delta_l^A(E) \right]. \tag{2}$$

where $p^2 = 2E$.

In order to obtain scattering phases upon a fullerene, the solution has to be found of the following equation

$$\left( \frac{1}{2}\frac{d^2}{dr^2} + W_C(r) - V_C^{pol}(r) - \frac{l(l+1)}{2r^2} - E \right) P_{El}^C(r) = 0, \tag{3}$$

The phases $\delta_l^C(E)$ can be found from asymptotic for $P_{El}^C(r)$ similar to (2).

To obtain phases for the electron-endohedral scattering one has to solve a combined equation

$$\left( \frac{1}{2}\frac{d^2}{dr^2} + \frac{Z}{r} - \hat{V}_{HF}(r) + W_C(r) - V_C^{pol}(r) - \frac{l(l+1)}{2r^2} - E \right) P_{El}^{A@C_N}(r) = 0 \tag{4}$$

The phases $\delta_l^{A@C_N}(E)$ can be found from asymptotic for $P_{El}^{A@C_N}(r)$ similar to (2). The details how to obtain scattering phases numerically one can find in [5].

We know quite well how to find $\hat{V}_{HF}(r)$ and have made a choice of $W_C(r)$ shape. More difficult is to determine an accurate expression for $V_C^{pol}(r)$ that is in fact an energy dependent and non-local operator. We have an experience to determine it for atoms employing perturbation theory and limiting ourselves by second order perturbation theory in incoming and fullerenes electrons interaction.

---

[1] We employ the atomic system of units, with electron mass *m*, electron charge *e*, and Planck constant $\hbar$ equal to 1.

For such an approach fullerene is a much more complex object than an atom. On the other hand, it exist a rather simple expression for $V_C^{pol}(r)$ (see e.g. [2])

$$V_C^{pol}(r) = -\frac{\alpha_C}{2(r^2+b^2)^2}, \qquad (5)$$

where $\alpha_C$ is the static dipole polarizability of a fullerene that for $C_{60}$ and a number of other fullerenes is measured and/or calculated (see [6] and references therein); $b$ is a parameter of the order of the fullerenes radius $R$. This expression had a long history of applications in electron-atom scattering studies and demonstrated acceptable accuracy.

3. To perform calculations, we have to choose concrete values for the $C_{60}$ potentials. The potential $W_C(r)$ is represented by a potential well with the depth 0.52 and inner $R_1$ (outer $R_2$) radiuses equal to $R_1 = 5.26$ ($R_2 = 8.17$). Note that $R = (R_1 + R_2)/2$. Results of calculations for scattering phases are illustrated by the case of Ar@$C_{60}$ in Fig. 1 and 2. In Fig.1 we compare $s$, $p$, $d$, $f$ phases of scattering upon Ar and Ar@$C_{60}$ that are obtained neglecting $V_C^{pol}(r)$. We see that the potential $W_C(r)$ supports three bound states in the $e + C_{60}$ system, in $s$, $p$, $d$-channels, that is seen from the fact that $\delta_{s,p,d}^{C_{60}}(E \to 0) \to \pi$. For $s$ and $p$-phases the Ar contribution is much bigger than that of $C_{60}$. The reason is simple – the account of exchange of the incoming electron with atomic electrons with simultaneous neglect of exchange between incoming and fullerenes electrons explains this. So, $\delta_{s(p)}^{Ar}(E \to 0) \to 3(2)\pi$, while $\delta_{s(p)}^{C_{60}}(E \to 0) \to \pi$. Note that for *d* and *f*-phases the endohedral and fullerenes coincide within the accuracy of numeric calculations.

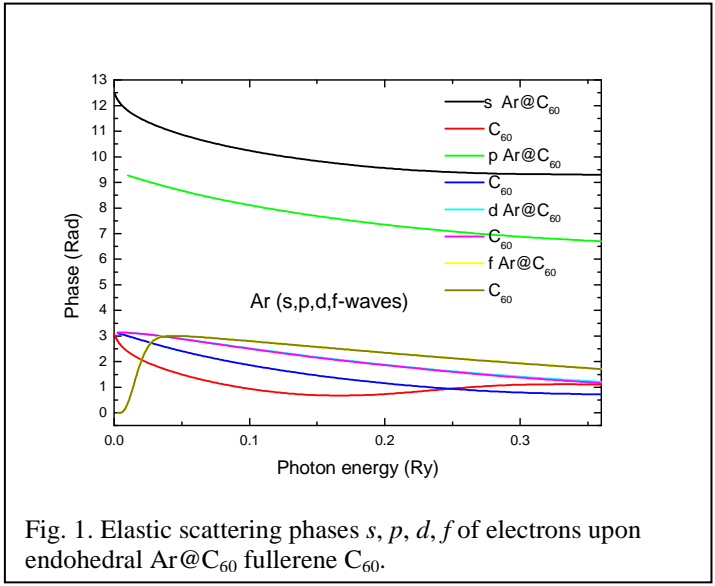

Fig. 1. Elastic scattering phases $s$, $p$, $d$, $f$ of electrons upon endohedral Ar@$C_{60}$ fullerene $C_{60}$.

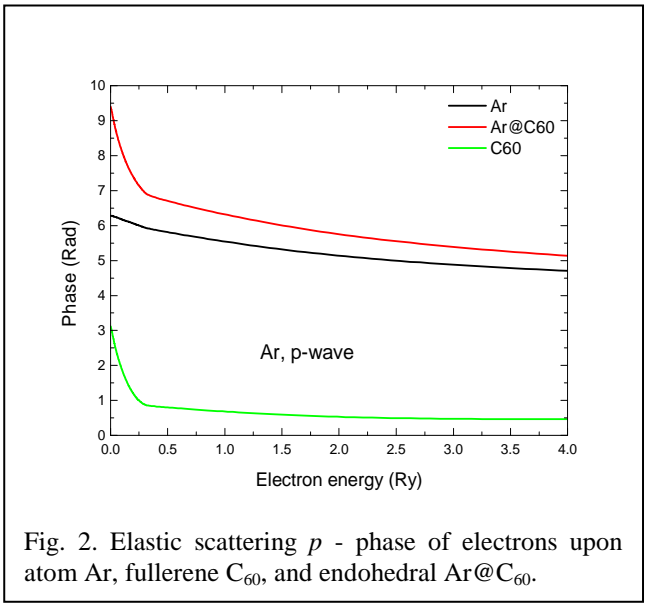

Fig. 2. Elastic scattering $p$ - phase of electrons upon atom Ar, fullerene $C_{60}$, and endohedral Ar@$C_{60}$.

Fig. 2 illustrates the *property of additivity*, namely the fact that the following equation is valid with quite high accuracy

$$\delta_l^{A@C_{60}}(E) = \delta_l^{A}(E) + \delta_l^{C_{60}}(E). \tag{6}$$

The calculation results are illustrated by the *p*-phase in collisions of electrons with Ar, $C_{60}$, and Ar@$C_{60}$. Qualitatively similar data are obtained for electron scattering upon Ne@$C_{60}$.

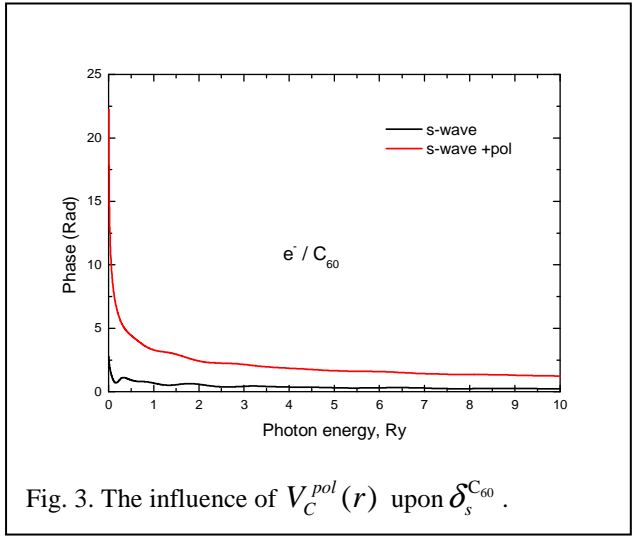

Fig. 3. The influence of $V_C^{pol}(r)$ upon $\delta_s^{C_{60}}$.

Inclusion of $V_C^{pol}(r)$ greatly modifies the phases, adding a prominent number of new bound states in the $e + C_{60}$ system. This is illustrated by Fig. 3, where we present *s*-phases in collisions of electrons with Ar@$C_{60}$ endohedrals.

If the electronic structure of the target, namely exchange between the incoming and target electrons will be taken into account, the phases will increase, acquiring an addition (in units of $\pi$) to the phase, equal to the number of bound in the target electrons with the given angular momentum *l*.

**4.** We have demonstrated using concrete examples that if the target is a loosely bound object, the rule of additivity (6) for scattering phases is valid. Derived for a concrete case, this rule is valid for any system since the physical reason for additivity is the ability to form bound states with each of the target atoms separately. On the other hand, if the target is tightly bound, the common phase can reach quite big values due to exchange between the incoming and target electrons. One can expect considerable changes of phases in collisions of a composite object due to addition to the target even of a single atom. This modification of phases leads to appearance of additional structure in the total elastic scattering cross sections.

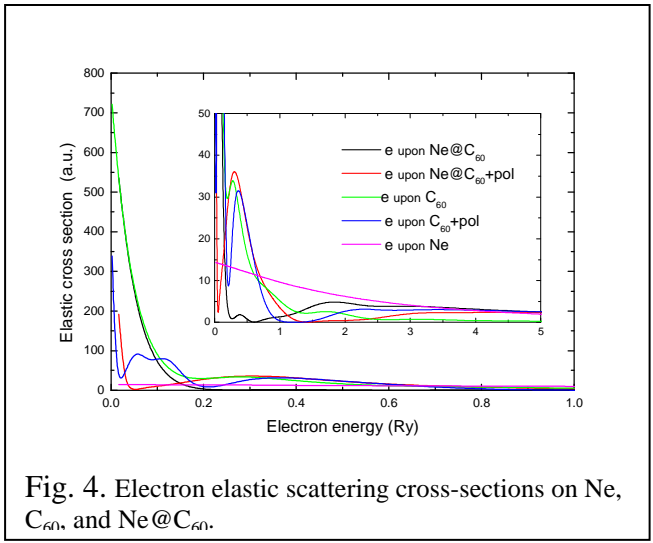

Fig. 4. Electron elastic scattering cross-sections on Ne, $C_{60}$, and Ne@$C_{60}$.

We illustrate it by Fig. 4, where cross-sections for scattering of electrons upon Ne, Ne@$C_{60}$ and $C_{60}$ are compared. For total cross-section more suitable is the Ne case. Here the role of correlations between atomic electrons is considerably less important than in Ar. At least, they do not modify essentially the low-energy cross-section. We see that the atomic cross-section at $E \to 0$ is much less than the cross-section upon $C_{60}$. Fullerene adds to the cross-section prominent structure. Its polarization potential strongly affects the cross-section. It is seen that phase additivity (6) does not lead to corresponding additivity in the cross-sections.

The presented results make it very interesting to study the elastic scattering of electrons by such multi-atomic systems as endohedrals. The modification of the cross-section due to presence even of a single additional atom can affect not only elastic scattering but some other connected to it properties and characteristics of electron-target collisions.

**5.** When the calculations presented in this Letter were finished, we became aware on similar results on *phase additivity*, obtained quite recently by V. K. Dolmatov and his students [7].

## References

[1]. M. Ya. Amusia, Chemical Physics **414**, 168–175 (2013)
[2]. L. D. Landau and E. M. Lifshits, *Quantum mechanics: non-relativistic theory*, 3$^{rd}$ edn, Pergamon Press, Oxford (1973)
[3]. M. Ya. Amusia, V. G. Yarzhemsky and L. V. Chernysheva, *Handbook of theoretical Atomic Physics, Data for photon absorption, electron scattering, vacancies decay*, Springer, Berlin, 2012, pp 806.
[4]. V. K. Dolmatov, J. L. King, and J. C. Oglesby, J. Phys. B **45**, 105102 (2012)
[5]. M. Ya. Amusia and L. V. Chernysheva, *Computation of atomic processes*, *A Handbook for the ATOM Programs*, Institute of Physics Publishing, Bristol and Philadelphia, 1997, pp.253
[6]. M. Ya. Amusia and A. S. Baltenkov, Phys. Let. A **360**, 294-298 (2006).
[7]. V. K. Dolmatov, *Private communication*, 16 February 2015.